# DIELECTRIC MIE VOIDS: CONFINING LIGHT IN AIR

M. Hentschel[1,*], K. Koshelev[2], F. Sterl[1], S. Both[1], J. Karst[1], L. Shamsafar[1],
T. Weiss[1,4], Y. Kivshar[2,*], and H. Giessen[1,*]

[1]*4th Physics Institute and Research Center SCoPE, University of Stuttgart, Pfaffenwaldring 57, 70569 Stuttgart, Germany*
[2]*Nonlinear Physics Centre, Research School of Physics, Australian National University, Canberra ACT 2601, Australia*
[3]*Institute of Physics, University of Graz, and NAWI Graz, Universitätsplatz 5, 8010 Graz, Austria*

[*]m.hentschel@physik.uni-stuttgart.de, yuri.kivshar@anu.edu.au, giessen@physik.uni-stuttgart.de

**Abstract**

Manipulating light on the nanoscale has become a central challenge in metadevices, resonant surfaces, nanoscale optical sensors, and many more, and it is largely based on resonant light confinement in dispersive and lossy metals and dielectrics. Here, we experimentally implement a novel strategy for dielectric nanophotonics: Resonant subwavelength confinement of light in air. We demonstrate that voids created in high-index dielectric host materials support localized resonant modes with exceptional optical properties. Due to the confinement in air, the modes do not suffer from the loss and dispersion of the dielectric host medium. We experimentally realize these *resonant Mie voids* by focused ion beam milling into bulk silicon wafers and experimentally demonstrate resonant light confinement down to the UV spectral range at 265 nm (4.68 eV). Furthermore, we utilize the bright, intense, and naturalistic colours for nanoscale colour printing. The combination of resonant dielectric Mie voids with dielectric nanoparticles will more than double the parameter space for the future design of metasurfaces and other micro- and nanoscale optical elements and push their operation into the blue and UV spectral range. In particular, this extension will enable novel antenna and structure designs which benefit from the full access to the modal field inside the void as well as the nearly free choice of the high-index material.

Resonant optical phenomena in metals and dielectrics have found profound applications in many fields[1]. The nanoscale confinement allows for unpreceded control of light-matter interaction at surfaces and interfaces[2], both in manipulating and controlling the flow of light[3-6] but also in enhancing the interaction with localized nanoscale systems such as emitters or sensor elements[7,8]. Resonant phenomena are usually associated with radiative and intrinsic loss channels, which are detrimental in many systems. In particular, metals show strong intrinsic losses. Thus, dielectric systems recently came into the focus of attention as they promised lower loss, higher degrees of flexibility with respect to tuning the interplay of different resonances,[9-12] as well as fabrication strategies closer to industrial standards. The optical response of these systems is best understood in terms of the well-known Mie theory[13], mostly related to the study of optical properties of high-index dielectric particles in nanophotonic systems[14-18], that form universal building blocks for optical metasurfaces[19-24] which are used in the manipulation[25-31], routing[32], and confinement of radiation[33], and thus trigger the start of the "Mie-tronic" era[34]. However, even in these systems, intrinsic loss plays an important role: The confinement takes place inside the high-index materials, most modal intensity is thus located inside the material – while this is of less concern in the near- and mid-infrared wavelength ranges, it becomes crucial for wavelength in the visible or even the UV spectral rage, where the intrinsic material loss hinders or even prevents the observation of resonant modes.

In this article we experimentally implement an elegant and powerful alternative route utilizing high-index materials, namely, the resonant confinement of light in air. In general, confinement in high-index dielectrics occurs due to a finite reflectance at the interface of the high-index material and air. For solid particles, the mode is thus localized within the high-index material. We demonstrate that also in the inverted case of an air void inside a homogenous high-index medium, localized optical modes emerge, which are confined within the nanoscale low-index void by virtue of the finite reflectance at the material discontinuity. Due to the confinement in air, the modes do not suffer from the loss and dispersion of the surrounding material. We show that these void modes are predicted by Mie's theory and bear close resemblance with the Mie modes of a high-index sphere, yet showing subtle, but significant differences.

We experimentally realize Mie voids by focused ion beam milling of cylindrical holes into bulk silicon wafers. Examples are shown in **Figure 1**. Panel a depicts a scanning electron microscope (SEM) image of a random arrangement of voids with varying diameter and depth. The optical microscope image (imaging setup: **panel f**) in panel **b** shows the distinctively different resonant optical scattering response of the voids, clearly depending *both* on diameter and depth. This underpins an additional strength of our FIB implementation: In contrast to the conventional fabrication method via reactive etching of silicon disks, we utilize the intrinsic depth variation capability as an additional degree of freedom. This feature is best visible for the chain of voids (panel **c** optical microscope image, panel **d:** top view SEM image). The focused ion beam cut (shown in panel **e**) demonstrates the significant differences in size *and* depth, manifesting themselves in a distinct color impression of each void (see microscope image in panel **c**). The dependence on diameter and depth clearly rules out trivial interference effects. The distinct colour of each void in the random as well as chain arrangement rules out grating phenomena, as grating colours would not vary across the grating and would only be determined by the grating constant (here 900 nm).

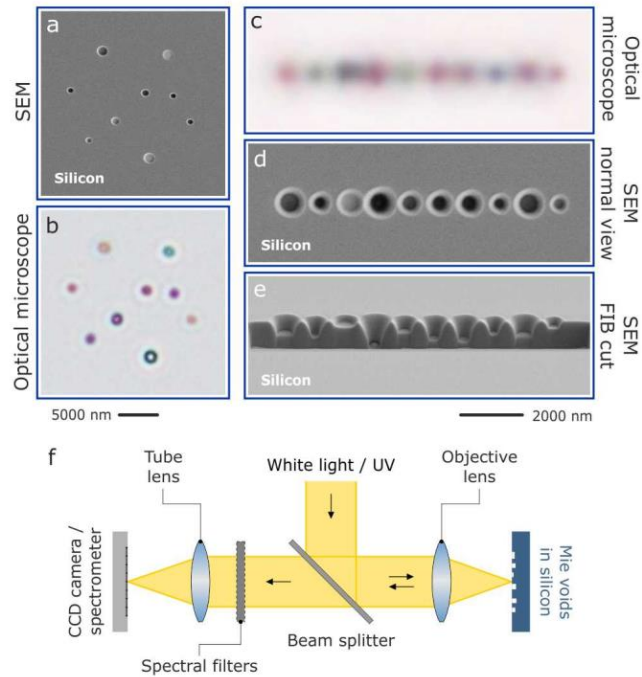

**Figure 1: Resonant dielectric Mie voids.** Focused ion beam milling allows to structure conically shaped voids of varying diameter and depth into a bulk silicon wafer. **a**, The scanning electron microscopy (SEM) image shows a random arrangement of holes of varying diameter and depth. **b**, In the optical microscope image one can observe the wavelength-dependent resonant scattering from the individual voids. Evidently, diameter and depth contribute to the resonant behaviour. The same behaviour is observed for the chain of voids with 900 nm distance with varying size and depth. **c**, Optical microscope image. **d**, Top view SEM image, **e**, SEM image focused ion beam cut. The distinct colour impression of each void can clearly be resolved in the optical microscope image. These experiments underpin the resonant and localized nature of the observed modes, which depend both on depth and diameter (or volume). **f**, Sketch of the experimental setups used. The surface of the silicon is illuminated with white light and the reflected light is collected. The setups allow to control the angle of incidence via apertures. Additionally, spectral filters (band pass filters) allow to image the structures at specific wavelength down into the UV spectral range.

As a first step we will demonstrate that Mie's theory describes the scattering of *voids* in a high-index surrounding, underpinning our experimental findings. In general, it is not possible to analytically solve Maxwell's equations for propagation and interaction of light for any given situation. Mie, however, was able to show that the scattering problem for a spherical inclusion of radius $R$ composed of a material with refractive index $n_i$ in a surrounding medium of refractive index $n_e$ can be solved analytically by a decomposition of the incident and outgoing waves into vector spherical harmonics. The strength of Mie's theory lies within these analytical solutions, which allow for intuitive insight into the behaviour of the system, e.g., the dependence on geometrical parameters or the complex dielectric function of the materials (which includes intrinsic loss). They also allow to deduce the near- and far-field behaviour. Interestingly, in the majority of cases, researchers have studied the case of high-index spheres in a low index surrounding, in particular for the implementation of high-index dielectric nanophotonics. Mie's solutions, however, describe a broader scenario as the model places no restrictions on the relative or absolute values of the dielectric function inside and outside the sphere, which opens a second, symmetric parameter space, namely low-index voids with refractive index $n_i$ placed inside a high-index dielectric host medium $n_e$ ($n_e > n_i$). While a similar phenomenon is known for metallic structures by virtue of Babinet's principle[35,36], in dielectric systems the full strength of this symmetry has not been explored or experimentally implemented. Optical resonances in textured, semi-periodic dielectric void systems of corrugated silicon surfaces have been studied and resonances associated with voids have been found, which were utilized for absorption enhancement[37] as well as light manipulation[38]. In nonporous, periodic gold surfaces void-based resonances were found to manipulate the diffraction of the surface, interpreting the voids as metallic Mie cavities[39]. Also, air bubbles in water have been studied with ray-optical as well as Mie scattering methods, however, no strong modal confinement can be observed as the refractive index contrast between the air bubbles and water is only on the order of 0.35. Additionally, bubble sizes are often in the several to hundreds of micron region (placing them into the realm of ray optics)[40,41]. Moreover, voids in host media have been investigated in the realm of effective material properties[42]. However, the confinement to the low-index void material and the fact that the surrounding host dielectric high-index material and its generally significant loss play a minute role for the properties of the modes has not been exploited and in particular not been experimentally implemented. It should be noted, that this behaviour is in stark contrast to inverse Babinet-type structures known in metallic systems, which are subjected to significant loss in the surrounding metal due to currents and fields inside the metal. Dielectric voids thus allow to push the resonant mode energies far into the visible and even into the UV spectral range, as we demonstrate below.

First, we notice that the denominator of the scattering amplitudes in Mie theory determines the eigenfrequencies of the resonant states, that is, the TM and TE modes in a spherical inclusion, obeying the following relations

$$\text{TM modes:} \quad \frac{\psi'_l(n_i k_0 R)}{\psi_l(n_i k_0 R)} = \frac{n_i}{n_e} \frac{\xi'_l(n_e k_0 R)}{\xi_l(n_e k_0 R)}, \quad (1)$$

$$\text{TE modes:} \quad \frac{\psi'_l(n_i k_0 R)}{\psi_l(n_i k_0 R)} = \frac{n_e}{n_i} \frac{\xi'_l(n_e k_0 R)}{\xi_l(n_e k_0 R)}. \quad (2)$$

The prime denotes derivatives with respect to the argument, while $k_0 = \omega/c$, and $\psi_l$ and $\xi_l$ are the Riccati-Bessel functions of order $l$, with $\psi_l(x) = x j_l(x)$ and $\xi_l(x) = x h_l(x)$, where $j_l$ and $h_l$ are the spherical Bessel and outgoing spherical Hankel functions, respectively. We assume non-magnetic materials ($\mu = 1, \varepsilon \neq 1, n = \sqrt{\varepsilon}$). Solving these relations, one can calculate the resonant wavelength, linewidth, and dispersion of the modes with material dispersion of refractive indices $n_e$ and $n_i$, and associated material losses, as well as the near- and far-field behaviour.

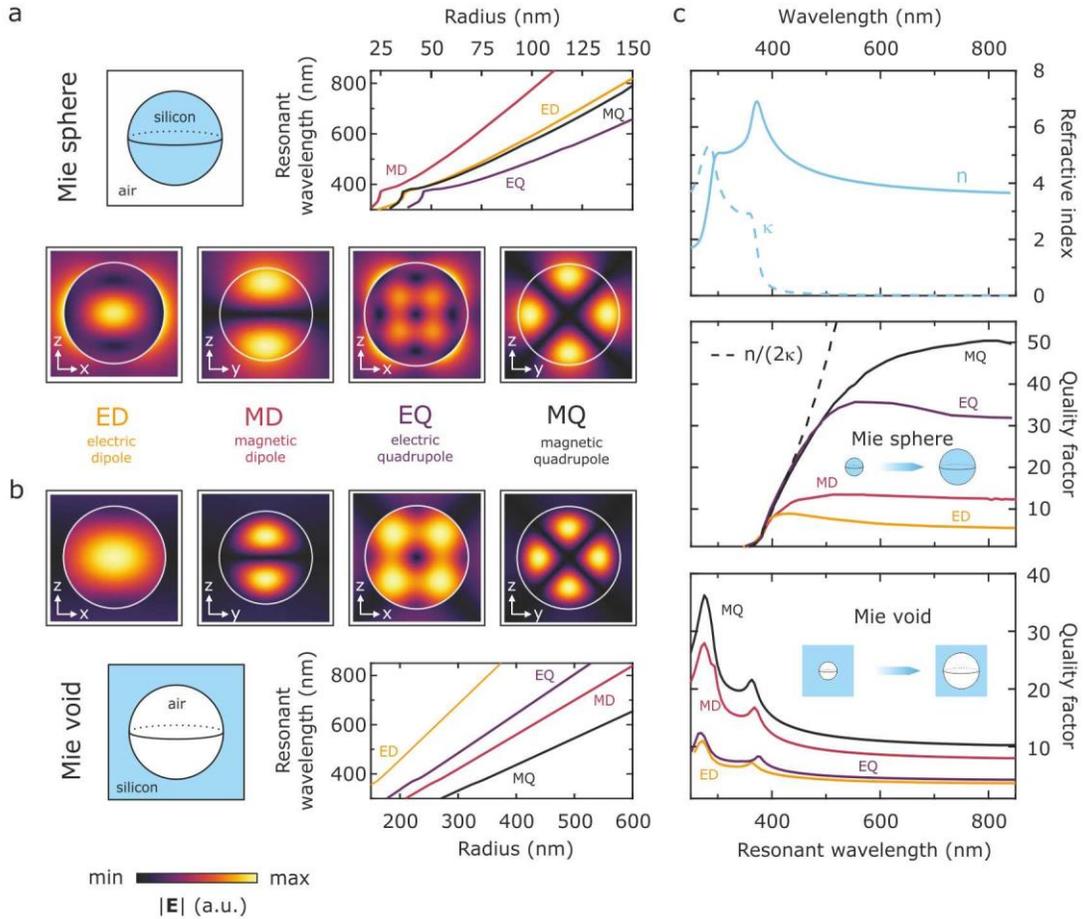

**Figure 2: Mie's theory for silicon particles in air and air voids in Si. a**, Field profiles and dispersion relations for localized Mie modes of a solid silicon Mie sphere in air and **b**, of a Mie void, which is, a spherical void inside an isotropic silicon environment. In both cases, Mie theory predicts the existence of localized modes (electric and magnetic dipole as well as quadrupole and higher-order modes). In spite of the inversed geometry, the modes show strong resemblance, yet, differ in the degree of confinement to the silicon and air, respectively. The dispersion relation of the modes of the silicon sphere is strongly influenced by the spectral dependence of the silicon refractive index and absorption (see **c**). For the Mie voids, in contrast, we observe perfectly linear dispersion, showing that the modes are fully localized in air without any noticeable influence of the silicon dispersion. **c**, Real part $n$ and imaginary part $\kappa$ of the refractive index of silicon (upper panel) in comparison to the properties of the sphere and void modes (middle and lower panel, respectively).[43] The graphs show the quality factors of each of the four fundamental modes for the sphere and void vs. the resonant wavelength of each of these modes. The dependence of the quality factor on the resonant wavelength is calculated by sweeping the sphere and void size. In case of the sphere (middle panel), one observes a drastic reduction of the quality factor for resonant wavelength approaching 400 nm and no modes

are observed for wavelengths shorter than 380 nm. This behaviour can be explained by the material losses in silicon, depicted with a dashed line showing the material quality factor $n/2k$. In case of the void (lower panel) the modes are confined within the air void. Consequently, the losses in silicon do not lead to a damping of the modes but rather result in a stronger confinement of the mode in the void by virtue of increased Fresnel reflection coefficient at the air-silicon interface (increased refractive index and increased absorption). Consequently, one can excite strongly confined modes in the deep-blue and UV spectral ranges with quality factors up to 40.

In **Figure 2** we display the analytical results for the eigenmodes and dispersion relations assuming silicon and air as the host and surrounding media, respectively. Silicon is a prime candidate for high-index dielectric nanophotonics due to its abundance, low cost, and well-established industry-level fabrication strategies [44,45]. **Figure 2a** depicts the analytical results for a silicon sphere in air. We calculate the four lowest energy eigenmodes and show one spatial cross section each. We were able to confirm the existence of the electric and magnetic dipolar modes as well as the electric and magnetic quadrupolar modes [12]. **Figure 2b** depicts the case of the dielectric Mie void. In spite of the very different geometry of sphere and void, the symmetry properties of both structures are rather similar. The analytical solutions of Mie sphere and Mie void show a significant resemblance of the near fields for all four fundamental modes. In spite of these similarities, there are subtle, yet important differences in the field distributions: In case of the silicon spheres, the modes are pinned to the silicon-air interface and most of the field is confined inside the high-index material, yet, also significant field components are extending into the lower index surrounding. In contrast, the Mie void modes show stronger confinement within the air void and barely any modal components extending into the silicon host. This is an intriguing phenomenon for two distinct reasons: The mode barely suffers from the intrinsic silicon loss while it simultaneously allows access to the entire modal field inside the air void. Contrary to the Mie sphere, one can thus push the resonant energies far into the visible and UV spectral range where silicon typically exhibits large absorption. We note that here we normalize the mode fields for Mie particles and voids identically using the exact analytic normalization procedure for open resonators[46].

Next, we compare the dispersion relations for the modes of Mie sphere and Mie void, which further underpin our interpretation. The graphs in **Figure 2a** and **2b** depict the radius dependence of the resonant wavelength of all four modes within the application-relevant wavelength regime from 250 nm to 850 nm. In case of the Mie sphere (**Figure 2a**) the dispersion relations are strongly influenced by the intrinsic silicon dielectric material dispersion, evidenced by their bending and the absorptive features around 380 nm. The Mie void modes exhibit perfectly linear dispersion (**Figure 2b**), indicating that the modes are indeed fully localized in air. Moreover, the Mie voids need to be significantly larger compared to the Mie sphere in order for the resonant wavelength of the modes to be in the same spectral range. The reason lies with the significant refractive index difference between silicon and air, which are on the order of $n_{\text{silicon}} \approx 4$ and $n_{\text{air}} \approx 1$. Moreover, we also observe a rearrangement of the energetic ordering of the modes. The fundamental eigenmode of the Mie sphere is the magnetic dipolar mode, whereas in case of the Mie void it is the electric dipolar mode.

The aforementioned results have emphasized the key role of the material dispersion on the optical and resonant properties. To further elucidate this behaviour, we plot in **Figure 2c** the real and imaginary part of the refractive index of silicon (upper panel) in comparison to the properties of the Mie sphere and void modes (middle and lower panel, respectively). The quality factors of each of the four fundamental modes are calculated and plotted against the resonant wavelength of each of these modes – the middle and lower two panels thus depict the quality factors of the modes for a radius sweep of sphere and void, respectively. In case of the sphere, one observes a drastic reduction of the quality factor for resonant wavelengths approaching 400 nm. More striking, no modes are observed for wavelengths shorter than ~ 380 nm. This behaviour can be explained by the optical loss in silicon. The dashed line indicates the material quality factor $n/(2\kappa)$, which is a measure for the relative contributions of dielectric response and material loss. As one can see, all modes in silicon are restricted by this quantity – no modes can be excited at frequencies beyond this line as intrinsic damping dominates over the resonant behaviour. In case of the voids, plotted in the lower panel, we observe a completely different situation. For longer wavelengths, the quality factors are on the order of 10, comparable to the ones observed for the dipolar Mie sphere modes. At shorter resonant wavelengths, one finds an increase in the quality factors, *in particular below the cut-off observed for the Mie sphere*. As the modes are confined in the air void, the loss in silicon does not lead to a damping of the modes. On the contrary, the increase in the refractive index (real part $n$) at ~380 nm and the increase in the loss/absorption (imaginary part $\kappa$) lead to a *stronger* confinement of the mode to the void by virtue of increased Fresnel reflection at the air-silicon interface (note that the peaks in the quality factor dispersion coincide with the material's resonant features in $n$ and $\kappa$ for silicon, displayed in the upper panel in **Figure 2c**). Consequently, the inverted geometry of the Mie void and the confinement to air instead of the high-index lossy material reveals its full strength: The structure supports strongly confined modes even in the UV spectral region with quality factors up to 40 without suffering from material loss. Similar behaviour has been observed in hollow-core photonic crystal fibres that can, theoretically, exhibit extremely low attenuation due to the guiding of light inside the air-filled core.[47–49] We observe a stronger confinement and thus larger quality factor of the Mie void modes for increasing refractive index and increasing loss as well as a full damping of the Mie sphere modes for increasing loss, underpinning the above interpretations. We also observe very weak dependence of Mie void mode resonant wavelengths on the complex dielectric function of the host material, which makes them robust to inevitable material parameter fluctuations due to fabrication tolerances.

We note that the conventional method of modal analysis by using scattering or extinction cross-section to evaluate mode wavelength, linewidth and quality factor are not applicable for the case of single Mie voids embedded in homogeneous and isotropic high-index media. The non-resonant background response due to high-index surroundings is strong and smears out the mode's resonant features in the spectrum of the Mie voids. To eliminate the effect of non-resonant oscillations for single voids in substrate, we consider evaluating the reflectance for silicon substrate with individual voids located in it. By proper normalization of the reflected signal to the bare response of the system without voids, the effect of background oscillations can be eliminated. Therefore, the reflectance spectra can be used for extraction of the mode's resonant properties, as shown below for both experimental and calculated data.

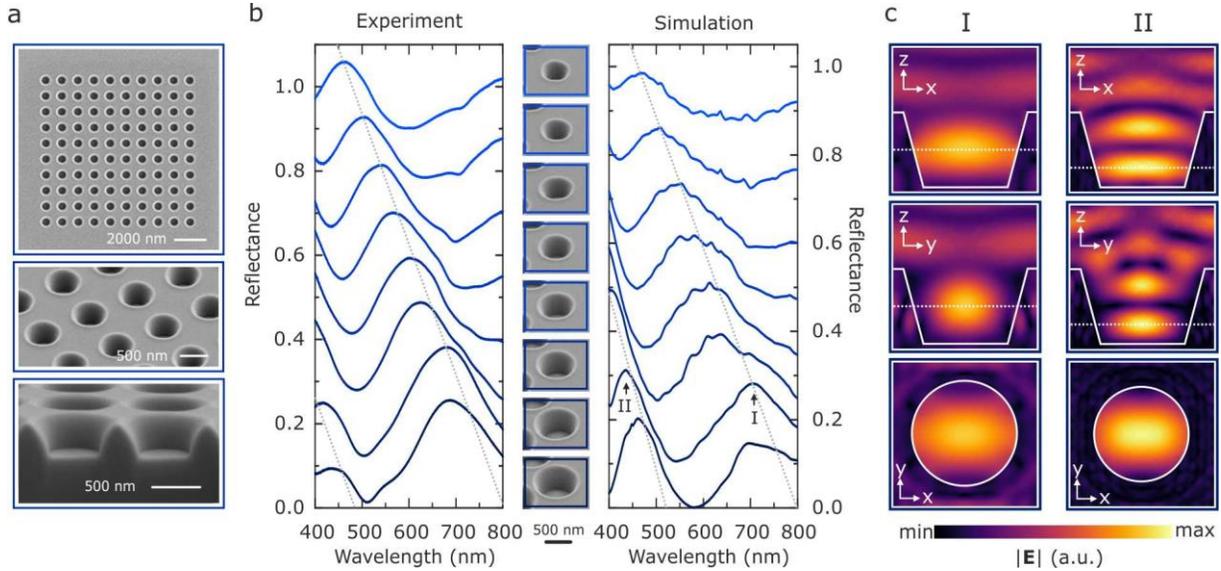

**Figure 3**: **Observation of Mie void modes**. **a**, SEM images of holes with a diameter of 610 nm and a depth of 410 nm arranged in a square grid of 900 nm periodicity in order to allow for spectroscopic measurement. The tilted view image of the same structures shows slightly inclined side walls, which are intrinsic to the milling process. The lowest image depicts a focused ion beam cut of holes with a 780 nm diameter viewed at a tilting angle of 80° underpinning the conical shape. **b**, Experimental and simulated normal incidence reflectance spectra of Mie voids in the silicon substrate. With increasing diameter, the fundamental mode I red-shifts while a higher-order mode II emerges, undergoing a spectral red-shift with increase of the diameter as well. **c**, Spatial distribution of the absolute value of the internal electric field for light incident along the Z axis and polarized along the X axis for the second largest Mie void. The observed field distributions at points I and II can be identified as the two fundamental modes predicted by Mie's theory in Figure 2, electric and magnetic dipole modes, respectively.

Our ansatz for an experimental realization is shown in more detail in **Figure 3**. Voids inside bulk silicon cannot be accessed optically in the visible wavelength range due to absorption in the host material. For an experimental implantation one needs to bring them close to the surface, as already shown in **Figure 1.** We utilize focused ion beam (FIB) milling into bulk silicon wafers to implement Mie voids as conically shaped voids at the silicon-air interface. **Figure 3a** depicts SEM images of our Mie voids, taken for the structures that are spectrally characterized. Reflectance spectra are measured from arrays of 10 x 10 elements at a period of 900 nm in both directions. The SEM images show voids with a diameter of ~610 nm and a depth of ~410 nm, illustrating the slightly inclined side walls, which are inherent to the milling process. The lowest image depicts a FIB cut of holes with 780 nm diameter at a tilting angle of 80° and underpins the conical shape.

**Figure 3b** shows the evolution of the experimental and simulated reflectance spectra of periodic void arrays with increasing diameter from ∼ 450 nm (uppermost spectrum) to ∼ 810 nm (lowest spectrum) with a depth of ∼ 450 nm. Arrangement in ordered arrays is done for the sake of simplicity of spectroscopic measurements while the resonant features originate from a single void. For the smallest void, we observe one resonant feature in the reflectance spectrum with strong localization in the void. This modal signature undergoes a spectral red-shift for increasing diameter. Additionally, a second feature appears at shorter wavelength, also undergoing a red-shift with increasing size. We note that these features come about due to the combination of the resonant scattering of the Mie voids and the large off-resonant reflectance of the bare silicon substrate on the order of 40%. The experimental spectra show excellent agreement with full-wave simulations of periodic arrays. One can see pronounced resonant features that confirm that reflectance analysis for single voids can be used to extract resonant properties from the spectrum along with the eigenmode simulations. The excellent agreement between experiment and simulation once more points to the localized nature of the observed modes. In order to elucidate the nature of the observed modes, we show electric field profiles for the second largest structure (see panel c) at wavelength points I and II at which we observe the modal signatures clearly in the field plots. The three cross-sections for both modes at the selected wavelengths are displayed in **Figure 3c**. Comparing the field distributions to the analytical Mie void modes depicted in **Figure 2b**, we observe significant resemblance. Despite the fact that the experimental Mie void is bound to the interface and has a truncated cylindrical shape rather than a spherical one, the simulated mode I shows excellent agreement with the fundamental electric-dipole Mie mode of the spherical void. For the higher-order mode II, we observe reasonable resemblance with the next higher-order magnetic-dipole mode of a spherical void. Here we show that the modes observed in the truncated conical voids can indeed be understood in terms of the leading order modes of a spherical Mie void. The calculation results show two additional benefits of Mie void modes over conventional Mie resonances of dielectric particles: The mode wavelength and field profile are robust to geometry change (from spherical to cylindrical and truncated conical shapes). Moreover, the void mode characteristics change weakly by going from an idealized void in a homogeneous environment to a realistic void in a substrate, which differs strikingly from the properties of Mie resonances of Si nanoparticles requiring a low-index substrate.

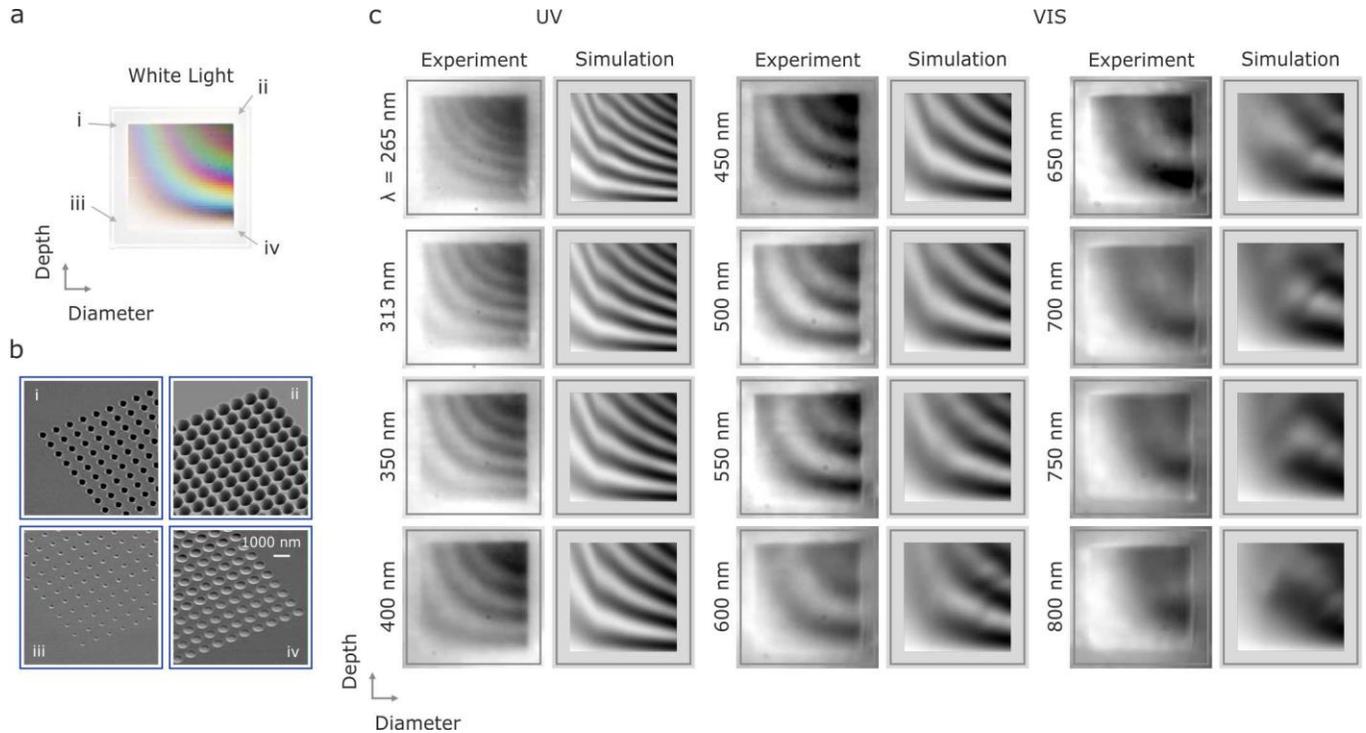

**Figure 4: Experimental observation of modes in the UV and visible spectral range. a**, White light reflection image of an array of dielectric voids in silicon. The size as well as the depth are varied in 36 steps each. The diameters range from ~330 nm to 750 nm and the depth from ~20 nm to 1100 nm. **b**, SEM images taken from the void array at the positions indicated in **a**. The images are taken for the smallest diameter voids for smallest and highest dose (shallowest (iii) and deepest voids (i), respectively) as well as for the largest void diameter for smallest and highest dose (shallowest (iv) and deepest (ii) voids, respectively). **c**, Experimental and simulated monochromatic images of the void array for the indicated wavelengths (displayed as vertical labels on the left). In the experiments, band pass filters at the specified centre wavelength are used to limit the wavelength range. With decreasing wavelength, an increasing number of modes can be observed. Due to the shorter wavelength, more and more higher-order modes can be resonantly excited. It is noteworthy, that this behaviour depends both on the depth as well as the diameter, ruling out trivial interference effects. In the UV spectral range at 265 nm or 4.7eV, one experimentally observes seven distinct higher-order modes. The simulated optical response is in good agreement with the experiment.

The results shown in **Figure 2** predict resonant behaviour down to the UV spectral range. In order to study the optical response of the Mie voids over a large range of sizes and depths and thus cover the spectral range from the UV to the near infrared we utilize an elegant imaging technique which allows to capture a large set of information at once. A fine variation of the geometrical parameters will enable the simultaneous observation of several mode orders when imaging the optical response at individual wavelengths. Depending on the diameter and the depth, the voids will support modes of different order, which all have the same wavelength. The white light microscope image of such a sample for "geometrical spectroscopy" is displayed in **Figure 4a**, while panel **b** shows selected SEM images. We vary the diameter and depth of the voids in 36 steps each, with diameters between ~330 nm to 750 nm and depths between ~20 nm and 1100 nm. The SEM images in panel **b** illustrate the extreme cases of these parameters at the four corners of the array: From the smallest and shallowest voids to the largest and deepest ones. Each "pixel" consists of 4 by 4 voids at a period of 900 nm and each pixel is spaced at 3,6 µm. The white light microscope image in panel **a** exhibits smooth and steady transitions in the colour impression of the structures, underpinning the fine parameter variation. One can clearly observe resonant mode scattering light at distinct wavelengths. The scattering behaviour depends both on the depth and diameter, ruling out a trivial interference phenomenon. Additionally, one can identify regions of identical or similar colour impression, showing that modes of different order are observed, which each scatter at a similar wavelength. This behaviour is particularly well visible for the blue and red stripes.

The white light impression already points to the measurement strategy: In order to identify modes of different order at the same wavelength, we image at distinct wavelength, separating the optical response of each "colour". We utilize band pass filters between 265 nm and 800 nm and take monochromatic images of the array, as shown in **Figure 4c**. These images thus directly visualize the resonant scattering behaviour of the structures at the respective wavelength. As mentioned, due to the variation of diameter and depth we directly visualize the dispersion of the modes with respect to these two parameters. Each experimental image is accompanied by a simulated one. The simulated optical response is based on the experimentally determined diameters and depth of the voids, *without any further assumptions*. In each image we observe strong modulation in the recorded far-field behaviour, which are caused by resonant scattering of the Mie voids, modifying the off-resonant reflectance of the bare silicon. Each modulation is thus associated to a resonant mode. With decreasing wavelength an increasing number of modulations and thus modes can be observed: For the image at 800 nm we only observe one mode, for 550 nm three modes can be identified, and for the shortest wavelength at 265 nm or 4.7 eV we identify seven modes. Due to the shorter wavelength an increasing number of higher-order modes can be resonantly excited. It is noteworthy that this behaviour depends both on the depth as well as the diameter. Overall, the simulations are in good agreement with the experiment.

As the *single* Mie voids displayed in **Figure 1b** show brilliant and naturalistic colours in their scattering response, our Mie void concepts holds great promise for nanoscale colour generation and colour printing [50,51]. We explore this application in **Figure 5**. Nanoscale colour printing that utilizes resonant scatterers depends on several key features, which are the quality factor, the scattering amplitude, and the tunability.[52–54] The resonances modify the reflected incident white light and are thus responsible for the colour impression. If the resonances are spectrally too small, the colours appear artificial and neon-coloured; if they are too broad or the scattering amplitude is too small, the colours are dull. As the image in **Figure 1b** as well as the spectra in **Figure 3c** clearly show, our Mie voids have ideal properties for brilliant and vivid colour generation in reflection. **Figure 5a** depicts a selection of our "colour cataloque" which has been used to generate our Mie void colour prints. For illustration, we show the colour impression of arrays of 10x10 elements with 900 nm period as well as SEM images for two cases: In the upper row voids of constant diameter and increasing depth, in the lower row for constant depth and increasing diameter. Both parameters allow to tune the colour impression over the entire visible wavelength range. We find that similar colours can be obtained from different combinations of diameter and depth, while the volume of the void determines the absolute scattering efficiency and thus the saturation of the colour.

As an example we choose the painting "Improvisation No. 9" painted in 1910 by Russian artist Wassily Kandinsky from the collection of the Staatsgalerie Stuttgart, Germany. **Figure 5b** depicts a section of the original painting, its experimentally colour-printed version, as well as normal and tilted view SEM images of the constituent Mie voids. The period is kept constant at 900 nm, ruling out any grating effect. Each pixel is thus composed *of one Mie void each*. As a consequence, even in the SEM image the contours and different components of the painting can be clearly identified. Additionally, it is obvious that the diameter and depth of the Mie voids serve as colour tuning knobs. Comparing the original painting and its colour printed counterpart, we find that the colour reproduction of the Mie-void colour printing is excellent, which is aided by the availability of "white" and "black" hues, that enable us to achieve large dynamic range and good contrast.

**Figure 5c** finally depicts the colour print of the entire painting "Improvisation No. 9" next to two human hairs for size comparison. The image has a size of 200 pixels by 200 pixels, a side length of 180 µm, and thus a resolution of 36.000 dpi. The colours are vivid and naturalistic and stand out against the unstructured bare silicon background of the wafer. As the colours are generated by voids in a silicon surface, the structure is remarkably long time stable due to the inertness and hardness of silicon.

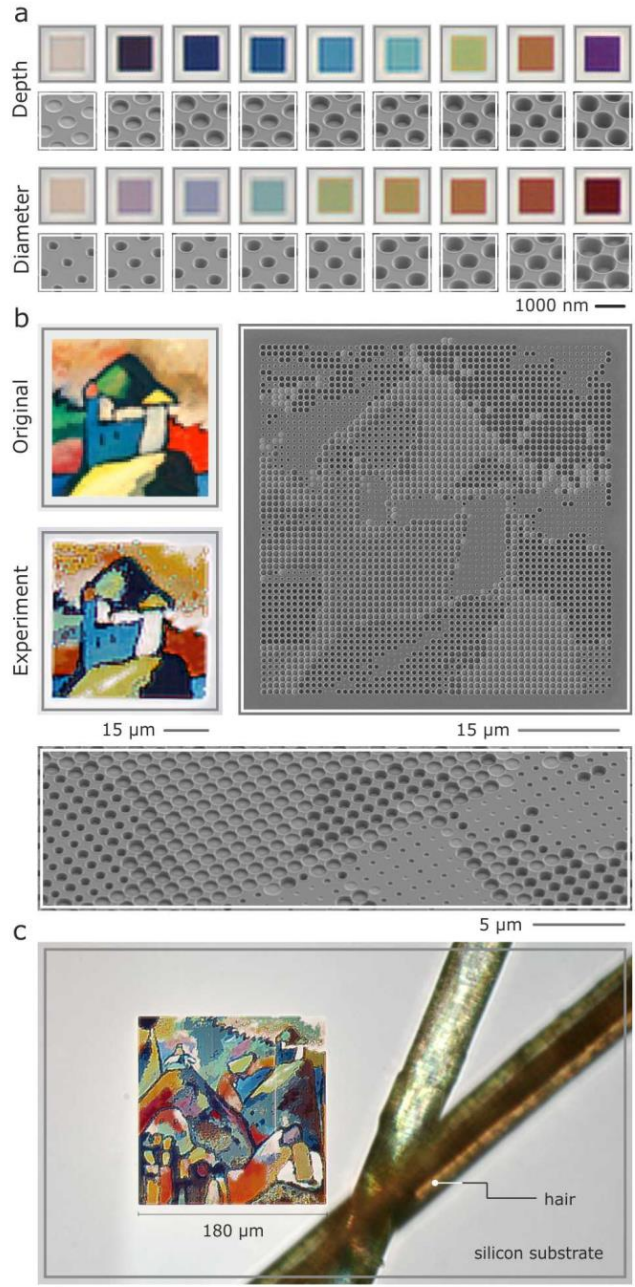

**Figure 5: Colour printing using Mie voids in a silicon substrate at 36.000 dpi**. Due to the ideal quality factor and thus line width of the Mie void modes, the structures show brilliant and naturalistic colours in reflection. Aided by the large interaction cross section, *one Mie void is sufficient for one pixel*. This gives our printing method a pointillistic touch, being completely different from grating-based approaches. **a**, Selected sizes and depth of Mie voids and their colour impression in an optical microscope. **b**, Detail taken from the painting "Improvisation No. 9" by Wassily Kandinsky (Staatsgalerie Stuttgart). The top left depicts the original artwork while the lower left shows an optical microscope image of the colour printed image. In the SEM image on the right, one can clearly identify the image as the pixel size is unchanged. In order to gain access to the full colour space, the diameter as well as the depth of the Mie voids has been varied, which is particularly well visible in the tilted SEM

image. **c**, Fully reproduced "Improvisation No. 9" in a silicon substrate. The image consists of 200 pixels x 200 pixels and has a side length of 180 μm, corresponding to 36.000 dpi resolution. The image is shown next to a human hair for size comparison.

In summary, we have introduced the concept of dielectric Mie voids. We have shown that Mie's theory predicts localized optical modes for low-index voids in a high-index surrounding, with the main benefit in a negligible influence of the surrounding high-index material and its high loss on the properties of the modes. In striking difference to conventional all-dielectric nanoparticles, Mie-void mode properties are very weakly dependent on the void geometry and substrate parameters. Moreover, they do not require any periodic arrangement for creating high-Q resonances. This allows to push the field of high-index nanophotonics into the blue and possibly even UV spectral range, significantly expanding the design parameter space for dielectric nano- and microoptical elements for the future design of metadevices.

Further experimental work in this direction can make use of other high-index materials, the combination of Mie spheres and Mie voids, as well as the combination of metallic plasmonic systems with dielectric ones, in the visible and UV regions. Mie voids are also particularly well suited for optical sensing[55] as well as trapping experiments and can utilize chiral structures.[56,57] Mie voids could also be used in hybrid systems where quantum emitters are coupled to the Mie voids which act as local nanoantennas. This should work particularly well in the blue and UV range, where many electronic resonances are located. Filling of a quantum emitter into the voids as well as using emitters in the void's host medium can be envisioned, for example excitons in gallium nitride or gallium arsenide, or defect centers in diamond or silicon carbide. Mie voids also hold great promise for reprogrammable structures,[58] switching, and active manipulation due to the possibly to fill the mode volume in the void with polymers or dielectrics. Moreover, Mie void modes are also expected to contribute to the optical, electronic, or acoustic properties in porous media, foams, glasses, dielectric and metallic networks.[59]